\documentclass[twocolumn,showpacs,prc,amsmath,amssymb,superscriptaddress]{revtex4}
\usepackage{graphicx}% Include figure files
\usepackage{dcolumn}% Align table columns on decimal point
\usepackage{bm}% bold math
\usepackage{epsfig}
\usepackage{fancyhdr}
\usepackage{multirow}
\usepackage{amsmath}
\usepackage{hyphenat}
\usepackage{amsfonts}
\usepackage{booktabs} % Para embellecer tablas
\begin{document}

\title{Proton induced fission of $^{181}$Ta at relativistic energies}

\author{Y.~Ayyad} \email{francescyassid.ayyad@usc.es} \affiliation{Universidade de Santiago de Compostela, E-15758 Santiago de Compostela Spain}
\author{J.~Benlliure} \affiliation{Universidade de Santiago de Compostela, E-15758 Santiago de Compostela Spain}
\author{E.~Casarejos} \affiliation{Universidade de Santiago de Compostela, E-15758 Santiago de Compostela Spain}
\author{H.~\'Alvarez-Pol} \affiliation{Universidade de Santiago de Compostela, E-15758 Santiago de Compostela Spain}
\author{A.~Bacquias} \affiliation{GSI, Planckstrasse 1, D-64941, Darmstadt, Germany}
\author{A.~Boudard} \affiliation{CEA-Saclay/IRFU, F-91191 Gif-sur-Ivette, France}
\author{M.~Caama\~no} \affiliation{Universidade de Santiago de Compostela, E-15758 Santiago de Compostela Spain}
\author{T.~Enqvist} \affiliation{GSI, Planckstrasse 1, D-64941, Darmstadt, Germany}
\author{V.~F\"ohr} \affiliation{GSI, Planckstrasse 1, D-64941, Darmstadt, Germany}
\author{A.~Keli\'{c}-Heil} \affiliation{GSI, Planckstrasse 1, D-64941, Darmstadt, Germany}
\author{K.~Kezzar} \affiliation{CEA-Saclay/IRFU, F-91191 Gif-sur-Ivette, France}
\author{S.~Leray} \affiliation{CEA-Saclay/IRFU, F-91191 Gif-sur-Ivette, France}
\author{C.~Paradela} \affiliation{Universidade de Santiago de Compostela, E-15758 Santiago de Compostela Spain}
\author{D.~P\'erez-Loureiro} \affiliation{Universidade de Santiago de Compostela, E-15758 Santiago de Compostela Spain}
\author{R.~Pleska\v{c}} \affiliation{GSI, Planckstrasse 1, D-64941, Darmstadt, Germany}
\author{D.~Tarr\'io} \affiliation{Universidade de Santiago de Compostela, E-15758 Santiago de Compostela Spain}

\date{\today}% It is always \today, today,
             %  but any date may be explicitly specified

\begin{abstract}

Total fission cross sections of $^{181}$Ta induced by protons at different relativistic energies have been measured at GSI, Darmstadt. The inverse kinematics technique used together with a dedicated set-up, made it possible to determine these cross sections with high accuracy. The new data obtained in this experiment will contribute to the understanding of the fission process at high excitation energies. The results are compared with data from previous experiments and systematics for proton-induced fission cross sections.

\end{abstract}

% PACS, the Physics and Astronomy
                             % Classification Scheme.
\par

%Use showkeys class option if keyword display desired

\maketitle

\section{\label{sec:level1}Introduction}

Spallation reactions induced by relativistic protons on $^{181}$Ta lead to excited target remnants with large fission barriers (20-25 MeV) covering a broad range in excitation energy. The investigation of the fission process under these extreme conditions is expected to provide relevant information on the dynamics of fission at high excitation energies. On the other hand, tantalum and tungsten alloys are proposed as optimum materials for the construction of spallation neutron sources because of their properties under extreme irradiation conditions: relatively large neutron production, corrosion resistance, and a high melting point. Spallation targets are of interest in different domains. One of the technologies which relies on spallation reactions is that of accelerator-driven systems, ADS~\cite{Nif01}, which are currently under study as an option for nuclear waste incineration. Recently the construction of the ESS (European Spallation Source)~\cite{Cla03} facility has been approved and different research communities are awaiting its opening to undertake a wide range of experimental programs in material science, biology and other scientific disciplines. Tantalum targets are also used for the production of exotic nuclei at ISOL-type~\cite{isolde} facilities and neutrinos~\cite{Bur96}. 

Fission may have a significant effect on the performance of a spallation target. Therefore, a good knowledge of the interaction of protons (commonly used as spallation sources drivers) with these materials is mandatory for their characterization. Reactions leading to fission are of interest because they contribute to the production of hazardous remnants, in particular gaseous ones, such as the isotopes of Kr and Xe. The composition of that radioactive inventory, its evolution, the influence of these changes to the target performance itself and its structural damages can be estimated with state-of-the-art models. However, only through an evaluation of numerical calculations with accurate data it is possible to validate these models and improve their reliability for use in technical applications.   

Nuclear fission is, indeed, a process which demands a complex description of the fissioning nucleus according to its excitation energy, angular momentum, fission barrier and the time taken to reach the scission point. A first approach to the fission description in terms of process probability was proposed by Bohr and Wheeler~\cite{Bohr39} from a purely statistical standpoint. In parallel, Kramers~\cite{Kra40} introduced a dynamical description of the fission process based on the coupling between internal (excitation energy) and collective (deformation) degrees of freedom through a dissipation parameter. Based on these ideas, Grang\'e {\it et al.}~\cite{Gra83} went a step further by including a time-dependent solution of the fission width that recently has been analytically formulated~\cite{Jur03,Jur05}. Many experimental results have already indicated the role of dissipative and transient effects in the fission process at high excitation energy~\cite{Jur04,Jac09}. Manifestations of these effects have also been observed in spallation-induced fission reactions~\cite{Ben06}, in particular with sub-actinide targets~\cite{Ben02}. 

Unfortunately, presently available data related to total fission cross sections of $^{181}$Ta above 700~MeV proton-beam energy are scarce and show clear discrepancies at 1000~MeV~\cite{Yur05,Boc78}. The situation does not improve at lower energies, where the available data are more abundant, but they also present inconsistent results~\cite{Bar62,Kon66, Shi73,Mau65}, in particular between 300 and 500~MeV. Most of these experiments were performed using passive track detectors and only few of them are based on coincident measurements of both fission fragments~\cite{Ste67}. Under such conditions it seems difficult to unambiguously identify a fission channel with a few mb cross section as is expected in this case. 

All previous measurements of fission reactions induced by protons on $^{181}$Ta were performed using the direct kinematics technique. Therefore, the reaction products have a very low kinetic energy, preventing their escape from the target. To overcome this difficulty the inverse kinematics technique was utilised in the present work. Several experiments performed at the FRS (FRagment Separator) spectrometer using the inverse kinematics technique~\cite{Fer05,Ben02,Ber03,Enq01,Ben01,Per07} measured the mass and charge of fission residues with high precision, in addition to the fission cross sections. However, the transmission of the fission fragments was limited by the acceptance of the spectrometer, and only one fission fragment was measured. Therefore, a dedicated experimental setup was used in the present work in order to register both fission fragments in coincidence with high efficiency and resolution~\cite{Sch00}.

In this work, we show the results of the experiment performed at GSI (Helmholtzzentrum f\"ur Schwerionenforschung) aimed at measuring the total fission cross section of $^{181}$Ta induced by protons in the energy range between 300 MeV and 1000 MeV taking advantage of the inverse kinematics. The dedicated experimental setup, enabled the measurement of the cross-sections with high precision. Our results will allow for benchmarking the state-of-the-art models, and provide data for systematic descriptions.

\section{\label{sec:level2} Experimental setup}

In the present experiment, the $^{181}$Ta nuclei were accelerated using the basic installation of GSI, the UNILAC linear accelerator and the SIS-18 synchrotron, up to 300, 500, 800 and 1000 A MeV with an intensity of the order of 10$^4$~ions/s and a spill duration of 7~s. These beams impinged onto a liquid hydrogen target. Due to the kinematics of the reaction, we were able to detect efficiently both fission fragments which were emitted in the forward direction with large kinetic energies. This capability enabled the use of relatively thick targets, increasing the statistics.

\begin{figure}[h]
\includegraphics[width=0.5\textwidth]{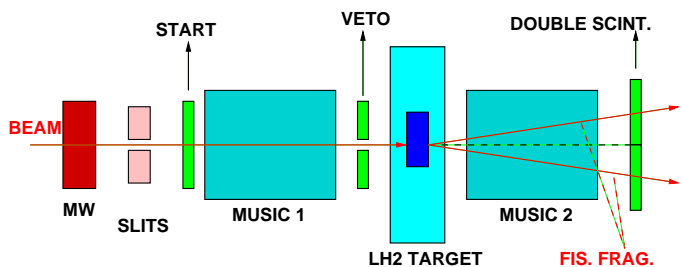}
\caption{\label{fig:setup} Schematic representation of the experimental setup used in the present experiment.}
\end{figure}

A sketch of the setup is shown in Fig.~\ref{fig:setup}. A Multi-wire chamber (MW) and thick iron slits were used to collimate the $^{181}$Ta beam at the target. A first scintillator detector (start) placed upstream of the target, determined the beam flux. The target consisted of a liquid hydrogen cell (85~mg/cm$^2$) inside a container with 100~$\mu$m titanium windows and a cryostat for liquefying the hydrogen. The target was surrounded by two Multi-sampling Ionization Chambers (MUSIC)~\cite{Pfu94} (200~mm x 80~mm window surface and 460~mm of active length) which measured the energy loss of the tantalum beam particles and that of the products of the reaction, respectively. These ionization chambers, having almost 100~\% efficiency for the detection of relativistic heavy nuclei, were used to identify reactions of $^{181}$Ta produced in the hydrogen target and in any other layer of matter present in the beam line. A veto scintillator with a 15~mm diameter hole placed just before the target allowed the rejection of beam-halo particles and misaligned beam trajectories.

The two fission fragments were detected independently, but in temporal coincidence, by a double paddle scintillator placed downstream of the target (300~mm x 70~mm and 3~mm thickness each paddle). According to the setup geometry, two different triggers were used for data acquisition: The ``beam" trigger was provided by the plastic scintillator placed upstream of the target in anticoincidence with the signal of the veto scintillator. 

The ``reaction" trigger was produced by the coincidence between the ``beam" trigger and the time-coincident signals on both paddles of the double scintillator placed downstream the target. These two triggers provided the measurement of the beam flux together with the fission events. The average rates for the ``beam" and ``reaction" triggers were around $10^4$ and 700~triggers/s, respectively. The ``beam" trigger was downscaled to reduce the data acquisition dead time.

\section{\label{sec:level3} Data analysis}

The identification of the fission events was based on the amplitude of the signals recorded by the two MUSIC chambers surrounding the target and the amplitude of the signals provided by the two paddles of the double plastic scintillators located downstream of the target. With this information, we were able to isolate fission events from other reactions channels occurring in the hydrogen target.

\begin{figure}[h]
\includegraphics[width=0.5\textwidth]{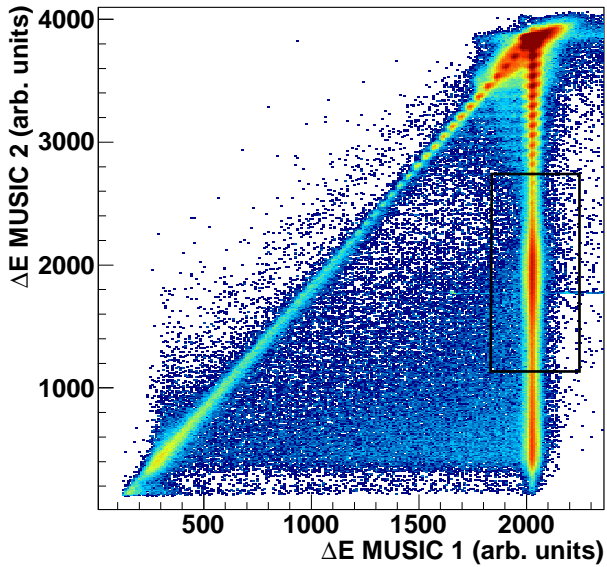}
\caption{\label{fig:music}Scatter plot of the amplitudes of the signals registered with the two MUSIC detectors. Nuclei lighter than $^{181}$Ta produced in reactions before MUSIC 1 appear in the diagonal region. In the vertical line the $^{181}$Ta spot corresponding to non-interacting beam particles is represented, and below the events corresponding to reactions in the target. The box encloses the fission region and the color code represents counts on the logarithmic scale.}
\end{figure}

In Fig.~\ref{fig:music} a scatter plot of the energy losses of ions traversing the two MUSICs, before and after the target, is depicted using the ``reaction" trigger. The events lying in the diagonal of this plot correspond to ions which kept their atomic number when passing through the target. These nuclei, lighter than the primary beam, have been produced in nuclear reactions induced by $^{181}$Ta projectiles in any layer of matter situated upstream of the hydrogen target. The dominant $^{181}$Ta spot of non-interacting beam particles is clearly visible at the top, near channel 3800 on the vertical axis. The vertical group below the beam spot corresponds to residual fragments produced in the interaction of $^{181}$Ta with hydrogen. In this group, events inducing high and small energy loss signals, correspond to residual heavy nuclei and emitted light nuclei from evaporation processes.

Since the energy loss of nuclei is proportional to their atomic number squared ({\it Z$^2$}), fission fragments are expected to produce energy loss signals corresponding to about half of the value obtained for the primary beam ({\it $\Delta E_{f.f.} \propto Z_1^{2} + Z_2^{2}= Z_{beam}^{2}/2$}). Therefore, fission products should be located around channel 1800 on the MUSIC 2 energy loss axis. To count for the fission events, {\it n$_{fiss}$}, a condition in the scatter plot shown in Fig.~\ref{fig:music} was applied selecting the region where the fission products are expected. Focusing on the selected region, indicated by the rectangular area in Fig.~\ref{fig:music}, the fission events were identified combining the amplitude (energy loss) of the signals recorded by the two paddles of the double plastic scintillator.

\begin{figure}[h!]
\includegraphics[width=0.5\textwidth]{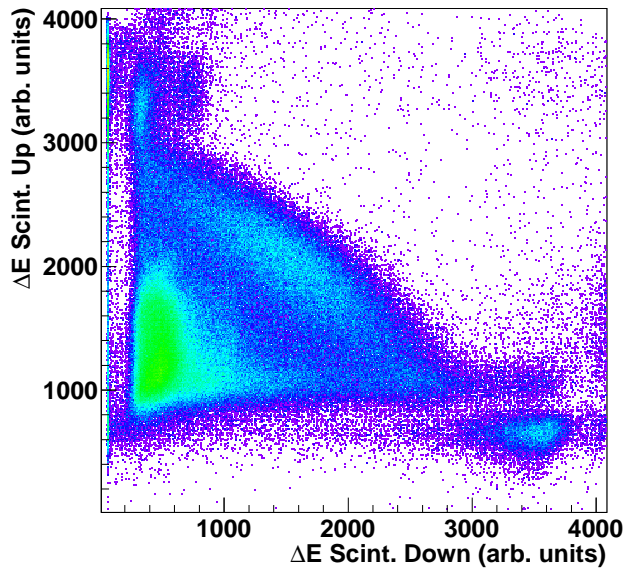}
\includegraphics[width=0.5\textwidth]{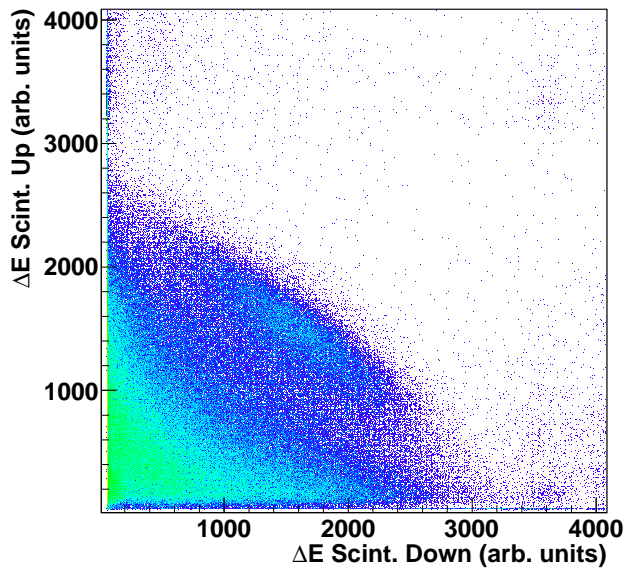}
\caption{\label{fig:scint}Scatter plot of the energy loss signals provided by the two paddles of the double plastic scintillator placed downstream the target with the reaction trigger (Upper panel 1~A GeV Full target - Lower panel 300~A MeV Full target). Both plots are normalized to the same number of counts to highlight the influence of different reaction channels. The color code represents counts on the logarithmic scale.}
\end{figure}

In Fig.~\ref{fig:scint} the amplitudes of the signals registered by both plastic scintillators in temporal coincidence (at 1~A GeV in upper panel and 300~A MeV lower panel), using the ``reaction" trigger, are represented in a scatter plot. Due to the charge splitting of the fission process, fission events  are expected to populate the diagonal band in this figure, and are separated from other much more abundant reaction channels. This fission region only represents a small part of the plot statistics since the fission probability is rather small. For this reason, fission events could only be properly identified by a detection setup enabling the identification of different reaction channels.

\begin{figure}[h!]
\includegraphics[width=0.5\textwidth]{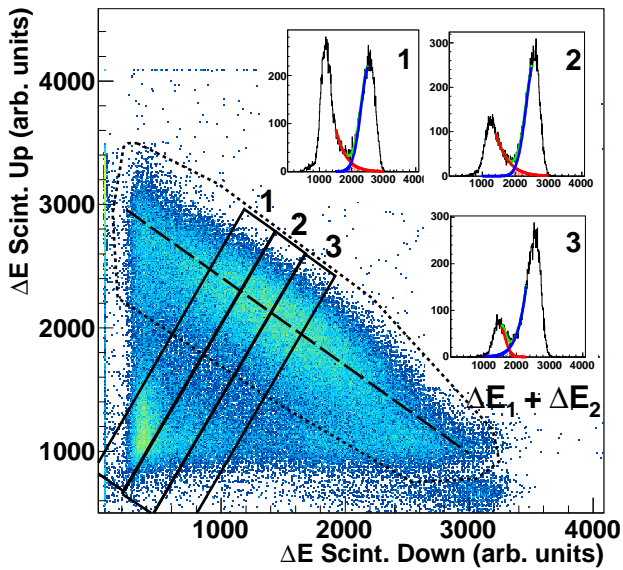}
\caption{\label{fig:cuts} Same as the upper panel in Fig.~\ref{fig:scint} but conditioned by the fission selection from Fig~\ref{fig:music}. The different contours and insets illustrate the background suppression method used to identify fission events as explained in the text.} 
\end{figure}

In order to provide an accurate measurement of the fission cross section, we evaluated the background which remains in the fission region due to simultaneous break-up and evaporation processes. To evaluate this background we used Fig.~\ref{fig:cuts}, where the energy loss provided by the two paddles for the double plastic scintillator at 1000~A MeV is represented by selecting only events compatible with a fission signal in the MUSIC detectors (rectangular area in Fig.~\ref{fig:music}). 

\begin{figure}[h!]
\includegraphics[width=0.5\textwidth]{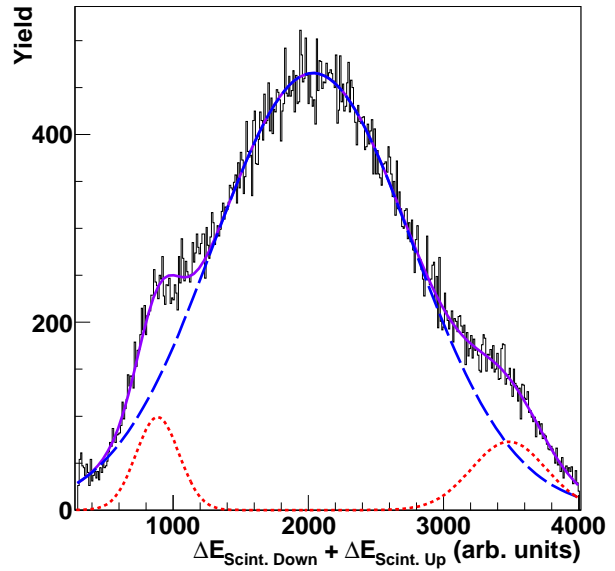}
\caption{\label{fig:evap} Projection of the fission region defined in Fig.~\ref{fig:cuts} on the {\it $\Delta E_1$+$\Delta E_2$} axis represented in the same figure. The contribution of evaporation residues was evaluated by means of a gaussian fit (dotted line) and subtracted from the total contribution (dashed line).} 
\end{figure}

In this figure, intermediate mass fragments (IMFs) produced in simultaneous break-up reactions may populate the fission region (dotted contour in Fig.~\ref{fig:cuts}). The evaluation of this break-up background was performed via dividing the fission region into slices as shown by the thin rectangles in Fig.~\ref{fig:cuts}. Each slice was then projected along its longitudinal dimension (insets in Fig.~\ref{fig:cuts}), which clearly enhanced the profile of the contributions coming from background  (left peak) and fission (right peak). Gaussian fits to each of the two contributions defined the correction for the break-up background suppression. On the other hand, evaporation residues could also populate the edges of the fission region  along an axis {\it $\Delta E_1$+$\Delta E_2$} (dashed line in Fig.~\ref{fig:cuts}) defined by the sum of the signals of the two scintillators. To overcome this problem, the region profile (dotted contour in Fig.~\ref{fig:cuts}) was projected onto this {\it $\Delta E_1$+$\Delta E_2$} axis to evaluate this contribution by means of gaussian fits as shown in Fig.~\ref{fig:evap}. 

The number of measured fission events {\it n$_{fiss}$} corresponds then to the number of events in the fission region corrected by the overlapped simultaneous break-up and evaporation background contributions.   

Fission yields (Y$_{fiss}$) were obtained from fission event measurements corrected by the background ({\it n$_{fiss}$}) and additional effects such as the secondary reactions of the fragments in the target ({\it f$_d$}) and the geometrical acceptance of the experimental setup ({\it f$_{geo}$}), according to the following equation:

\begin{eqnarray}
Y_{fiss}= n_{fiss} \cdot {f_d} \cdot f_{geo}.
\label{eq:yield1}
\end{eqnarray}

Secondary reactions of the fission fragments in the target were evaluated using the Karol's microscopic model~\cite{Kar75} and amounted to less than $2.5~\%$ for full target and less than $0.5~\%$ for empty target measurements at 1000~A MeV. Geometrical constraints were also considered to evaluate the efficiency of the detection setup. Fission products emitted close to the double paddle scintillator gap had a probability of passing through it or through the same paddle. A Monte Carlo simulation based on the liquid-drop-model calculation of the post-scission kinetic energy of the fission fragments~\cite{Wil76} was performed to evaluate the ratio of fission product losses due to the geometry of the setup. Taking into account the dispersion of the beam as measured with the Multi-wire chamber detector and the alignment, the coulomb force between both fragments and the distance from the centre of the hydrogen target to the double plastic scintillator, we calculated the perpendicular dimensions of the fission fragments distribution in the double plastic scintillator detection plane. The acceptance of these scintillators (300 mm x 140 mm and 1 mm gap) allowed us to evaluate the losses. Thus, the resulting yield was corrected by a geometrical factor ({\it f$_{geo}$}) having a value smaller than $10~\%$ at 1000 A MeV and decreasing with the beam energy. 

\begin{figure}[h]
\includegraphics[width=0.5\textwidth]{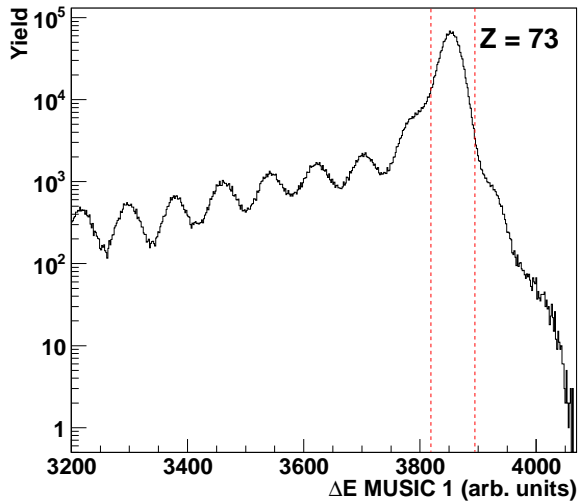}
\caption{\label{fig:music1} Energy Loss signals provided by the MUSIC 1. The region between the dotted red lines corresponds to $^{181}$Ta.}
\end{figure}

To determine the number of projectiles ({\it n$_{b}$}), we used the first MUSIC to identify tantalum among other nuclei that have been created in other layers of matter placed in the beam line before the target, as shown in Fig.~\ref{fig:music1}. The sum of the {\it Z}=73 ions identified according to this procedure using the ``beam" and the ``reaction" trigger corrected by the downscaling factor provided the total number of projectiles.

Due to the relatively large thickness of the target, the number of projectiles was corrected by a factor {\it f$_a$} ($< 5~\%$ for full target and $< 1~\%$ for empty target at 1000~A MeV) taking into account the attenuation of the beam intensity along the target. The value of this factor was also evaluated using Karol's model.

To correct for reactions taking place in the target windows (namely Ti) which surrounded the liquid hydrogen, fission yields determined with the empty target following the same analysis procedure were subtracted from the fission yield obtained with the full target. Finally, the respective fission yields were normalized to the number of projectiles and the number of nuclei in the target per unit area ({\it N$_t$}) to determine the total fission cross section according to the following expression (with {\it N$_{b}$ = n$_{b}$$\cdot$f$_a$}):

\begin{eqnarray}
\sigma=\left(\frac{Y_{fiss}^{full}}{N_{b}^{full}} -\frac{Y_{fiss}^{empty}}{N_{b}^{empty}}\right)\cdot \frac{1}{N_t}.
\label{eq:xs}
\end{eqnarray}

Particular attention was paid to the evaluation of the corresponding uncertainties. The main sources of systematic uncertainties were the identification of fission events ranging from 2 to 10\% ($\epsilon$($n_{fiss}$)), the beam intensity ($\approx 5\%$) and the target thickness ($\approx 4\%$). The sources of systematic uncertainty for the different correction factors were also evaluated. The systematic uncertainty of the geometrical correction factor ({\it f$_{geo}$}) was estimated to be smaller than 5~\%. The evaluation was done by changing the size of the double plastic scintillator gap and the beam profile in our simulation. The value of the systematic uncertainty of the correction factors due to the beam attenuation ({\it f$_a$}) and the secondary reactions of the fission fragments  ({\it f$_d$}) were smaller than $1~\%$ and almost the same for all energies. Due to the relatively large number of recorded fission events the statistical uncertainties were below 1.5~\%. Statistical and systematic uncertainties, others than the ones associated to the beam intensity and target thickness, for the measurements with the full and empty target are presented in Table~\ref{tab:table1} and Table~\ref{tab:table12}.

\begin{table}[h]
\caption{\label{tab:table1}Statistical ($\epsilon_{stat.}$) and systematic uncertainties due to the identification of fission fragments ($\epsilon$(n$_{fiss}$)), geometrical acceptance ($\epsilon$(f$_{geo}$)) and the attenuation ($\epsilon$(f$_a$)) of the beam affecting our measurements.}
\begin{ruledtabular}
\begin{tabular}{lccccc}

%\multicolumn{1}{c}{ Full Target} & \multicolumn{4}{c}{ Empty Target}
& \\
Energy &  $\epsilon_{stat.}$ & $\epsilon$(n$_{fiss}$) & $\epsilon$(f$_{geo}$)  & $\epsilon$(f$_a$) & $\epsilon$(f$_d$) \\
\hline
  & \\
1000~A MeV & 0.38\%  & 6.01\% & 4.32\% & 0.43\% & 0.22\%\\ 
800~A MeV & 0.39\% & 7.16\% & 3.72\% & 0.41\% & 0.22\%\\ 
500~A MeV & 0.45\% & 7.84\% & 2.80\% & 0.39\% & 0.22\%\\
300~A MeV & 0.33\% & 9.51\% & 2.05\% & 0.38\% & 0.22\%\\

\end{tabular}
\end{ruledtabular}
\end{table}

\begin{table}[h]
\caption{\label{tab:table12} Same as Table~\ref{tab:table1} but for empty target measurements. (*At 500~A MeV the independent number of fission could not be determined but the yield normalized to the number of projectiles.)}
\begin{ruledtabular}
\begin{tabular}{lccccc}

%& \multicolumn{2}{c}{ Full Target} & \multicolumn{2}{c}{ Empty Target}
& \\
Energy &  $\epsilon_{stat.}$  & $\epsilon$(n$_{fiss}$) & $\epsilon$(f$_{geo}$) & $\epsilon$(f$_a$) & $\epsilon$(f$_d$)\\
\hline
  & \\
1000~A MeV & 1.45\%  & 1.95\% & 4.32\% & 0.05\% & 0.04\%\\ 
800~A MeV & 1.05\% & 4.56\% & 3.72\% & 0.05\% & 0.04\%\\ 
500~A MeV* & - & 6.71\% & 2.80\% & 0.05\% & 0.04\%\\ 
300~A MeV & 0.96\% & 7.08\% & 2.05\% & 0.05\% & 0.04\%\\

\end{tabular}
\end{ruledtabular}
\end{table}

\section{\label{sec:level4}Results}

Using the method described in the previous sections, we have measured with high precision the total fission cross section of  $^{181}$Ta induced by protons at 300, 500, 800 and 1000~A MeV. The results obtained for each energy are presented in Table~\ref{tab:table2}. The magnitude of the measured cross sections is rather small and strongly decreases for the lower beam energies. The associated uncertainties are also rather small ($\approx 10~\%$) but increase for the lowest energies ($\approx 18~\%$) since the smaller fission cross sections complicates the identification of fission events.

\begin{table}[h]
\caption{\label{tab:table2} Total fission cross sections determined in this work}
\begin{ruledtabular}
\begin{tabular}{lcccc}
Energy & Fiss. cross section & Stat. uncert. & Syst. uncert. \\
(A MeV) & (mb) & (\%) & (\%)\\
\hline
1000 & 20.17~$\pm$~2.19 & 0.46 & 10.85 \\ 
800 & 13.09~$\pm$~1.62  & 0.32 & 12.34\\ 
500 & 7.53~$\pm$~1.40 & 0.51 & 18.54\\ 
300 & 6.55~$\pm$~1.00 & 0.48 & 15.21\\ 
\end{tabular}
\end{ruledtabular}
\end{table}

\begin{figure}[h]
\includegraphics[width=0.5\textwidth]{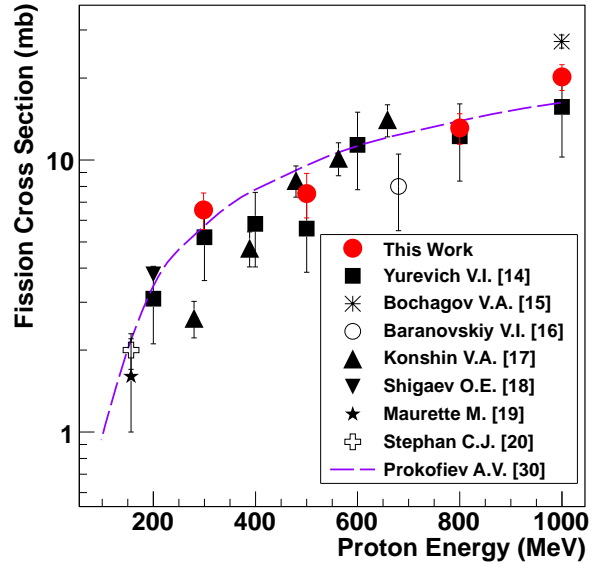}
\caption{\label{fig:data} Fission cross sections measured in this work (solid circles) in comparison to previously measured data and estimates obtained from the systematics established by Prokofiev (dashed line).}
\end{figure}

In Fig.~\ref{fig:data}, we present the cross sections obtained in this work as solid points compared to previous measurements by different authors. In this figure, we also present predictions obtained from the systematics established by Prokofiev some years ago~\cite{Pro01} (dashed line). 

From the analysis of the previously measured cross sections, one can identify some clear discrepancies. At the highest energies, one can find two rather discrepant measurements around 670~MeV by Konshin {\it et al.}~\cite{Kon66} (14.0~$\pm$~1.9~mb) and by Baranovskiy {\it et al.}~\cite{Bar62} (8.0~$\pm$~2.5~mb). At 800~MeV it exists a single measurement by Yurevich {\it et al.}~\cite{Yur05} and at 1000~MeV one finds again two discrepant values obtained by Yurevich {\it et al.} (15.65~$\pm$~5.4~mb) and Bochagov {\it et al.}~\cite{Boc78} (27.0~$\pm$~1.5~mb). Our results are in very good agreement with the measurement of Yurevich {\it et al.} at 800~MeV and within the error bars at 1000~MeV, solving the existing discrepancy in this energy range. Moreover, we also confirm the predictions estimated by the systematics of Prokofiev.

In the energy range between 300 and 600~MeV, we can also observe important discrepancies between different measurements. Around 300~MeV the data obtained by Yurevich {\it et al.}  (5.2~$\pm$~1.6~mb) and Konshin {\it et al.} (2.6~$\pm$~0.4 mb) differ by a factor two. Around 400~MeV, the measurements by Yurevich {\it et al.}  (5.79~$\pm$~1.78~mb) and Konshin {\it et al.} (4.7~$\pm$~0.7 mb) are in rather good agreement. However, the measurement by Konshin {\it et al.} is significantly smaller than the one obtained from the systematics of Prokofiev (7.60~mb). The measurement by Yurevich {\it et al.} could be compatible with the systematics owing to its large uncertainty. Finally, around 500~MeV the measurements by Konshin {\it et al.} (8.3$~\pm$~1.1~mb) and Yurevich {\it et al.} (5.59~$\pm$~1.72~mb) also differ by a large factor. Our measurements at 300 and 500~MeV are consistent with the estimated values from the Prokofiev formula, and, confirm the largest values of the cross sections measured in this region.
  
From this analysis we can conclude that our data confirm the measurements by Yurevich {\it et al.} and the cross sections estimated by the systematics above 700~MeV. At lower energies our measurements clarify the discrepancies existing until now. In the energy range between 300 and 600~MeV, our data favor those measurements presenting the highest cross sections. Moreover our data confirm the predictions obtained by the systematics of Prokofiev over the entire energy range covered by this work.

\section{\label{sec:level5}Summary and conclusions}

In this work we have investigated the proton induced fission of  $^{181}$Ta in inverse kinematics at 300, 500, 800 and 1000~A MeV. The combination of the inverse kinematics technique with a highly efficient detection setup made it possible to determine the total fission cross sections with high accuracy. The coincident measurement between both fission fragments and their identification from the rough determination of their atomic number allowed to clearly identify and separate the fission events from other reaction channels. This selection is shown to be extremely useful at lower energy when the fission cross section is small and complicates the identification of this reaction channel.

The new data have completed the scarce number of measurements existing above 700 MeV. At intermediate energies, the quality of the new data enabled to clarify previous results. Moreover, these new data are overall in good agreement with the systematics established by Prokofiev over the entire energy range. The results will also allow to benchmark different state-of-the-art models describing fission and transport simulation codes used in the design of spallation targets.

\begin{acknowledgments}

This work was partially supported by the European Commission under project FI6W-CT-2004-516520 EUROTRANS, by the Spanish Ministry of Research and Innovation under the grant FPA2007-62652 and BES-2008-005553, the project consolider-CPAN CSD2007-00042 and the Regional Government of Galicia under the program ``grupos de referencia competitiva" 2006/46.

\end{acknowledgments}

%\bibliography{ta} % Produces the bibliography via BibTeX.

\end{document}